\documentclass[conference, final]{IEEEtran}
\usepackage{amsmath,amssymb,amsfonts}
\usepackage{graphicx}
\usepackage{textcomp}
\usepackage{xcolor}
\usepackage{physics}
\usepackage{booktabs}
\usepackage{braket}
\usepackage{microtype}
\usepackage{algorithm}
\usepackage[noend]{algpseudocode}
\usepackage{pdfpages}
\usepackage{tikz}
\usepackage[backend=biber, style=ieee, maxnames=5, minnames=1]{biblatex}
\usepackage{hyperref}
\graphicspath{ {./figures/} }

\newcommand{\bigO}{\mathcal{O}}

\newcommand\copyrighttext{%
  \footnotesize \textcopyright 2024 IEEE. Personal use of this material is permitted.
  Permission from IEEE must be obtained for all other uses, in any current or future
  media, including reprinting/republishing this material for advertising or promotional purposes, creating new collective works, for resale or redistribution to servers or lists, or reuse of any copyrighted component of this work in other works. 
  DOI: \href{https://doi.org/10.1109/QCE60285.2024.00169}{10.1109/QCE60285.2024.00169}
  }
\newcommand\copyrightnotice{%
\begin{tikzpicture}[remember picture,overlay]
\node[anchor=south,yshift=10pt] at (current page.south) {\fbox{\parbox{\dimexpr\textwidth-\fboxsep-\fboxrule\relax}{\copyrighttext}}};
\end{tikzpicture}%
}

\begin{document}

\title{QUACK: Quantum Aligned Centroid Kernel}

\author{\IEEEauthorblockN{Kilian Tscharke, Sebastian Issel, Pascal Debus}
\IEEEauthorblockA{\textit{Quantum Security Technologies} \\
\textit{Fraunhofer Institute for Applied and Integrated Security}\\
Garching near Munich, Germany \\
\textit{$\langle$firstname$\rangle$}.\textit{$\langle$lastname$\rangle$}@aisec.fraunhofer.de}
}

\maketitle
\copyrightnotice

\begin{abstract} 
Quantum computing (QC) seems to show potential for application in machine learning (ML). In particular quantum kernel methods (QKM) exhibit promising properties for use in supervised ML tasks. However, a major disadvantage of kernel methods is their unfavorable quadratic scaling with the number of training samples. Together with the limits imposed by currently available quantum hardware (NISQ devices) with their low qubit coherence times, small number of qubits, and high error rates, the use of QC in ML at an industrially relevant scale is currently impossible. As a small step in improving the potential applications of QKMs, we introduce QUACK, a quantum kernel algorithm whose time complexity scales linear with the number of samples during training, and independent of the number of training samples in the inference stage. In the training process, only the kernel entries for the samples and the centers of the classes are calculated, i.e. the maximum shape of the kernel for n samples and c classes is (n, c). During training, the parameters of the quantum kernel and the positions of the centroids are optimized iteratively. In the inference stage, for every new sample the circuit is only evaluated for every centroid, i.e. c times.
We show that the QUACK algorithm nevertheless provides satisfactory results and can perform at a similar level as classical kernel methods with quadratic scaling during training. In addition, our (simulated) algorithm is able to handle high-dimensional datasets such as MNIST with 784 features without any dimensionality reduction.
\end{abstract}


\section{Introduction}
Supervised Learning is an important branch of Machine Learning (ML) where a model is trained on labeled data to predict the labels of new, unseen data. It encompasses two main types of tasks: classification, which predicts discrete labels or classes, and regression, which forecasts continuous values.
Quantum Machine Learning (QML) is an emerging field in the intersection of Quantum Computing (QC) and ML with the goal of utilizing the potential advantages of QC - like superposition, entanglement, and the exponential size of the Hilbert space - for Machine Learning. In particular, Quantum Kernel Methods (QKM) have recently gained attention because of their ability to replace many supervised quantum models and their guarantee to find equally good or better quantum models than variational circuits \cite{supervised_qml_are_kernel_2021}. In addition, theoretical results are showing that QKMs can handle classification problems that cannot be solved using classical ML techniques, such as classifying numbers based on the discrete logarithm \cite{quantum_kernel_advantage_discrete_log_2021}.
However, a significant drawback of using (quantum) kernels is the quadratic time complexity of the kernel calculation, i.e. $\bigO(n_\text{train}^2)$ for $n_\text{train}$ training samples, since a kernel value must be estimated for each pair of samples. For the inference stage, the time complexity without using advanced techniques such as Support Vectors is $\bigO(n_{\text{train}} n_{\text{predict}})$. Estimating a quantum kernel for the - by classical ML standards very small - URL dataset \cite{URL_dataset_2016} with around 36,000 samples, requires approximately $10^{9}$ kernel value calculations. With the commonly-used number of 1,000 shots, this involves $10^{12}$ circuit executions. A state-of-the-art IBM device with Eagle r3 processor (e.g. \textit{ibm\_sherbrooke} \cite{sherbrooke}) achieves 5,000 CLOPS (circuit executions per second) and hence the execution time for calculating the kernel would be $10^{8}$ seconds, or over 6 years. 

Quantum Kernel Alignment (QKA) is a fascinating tool for QKMs that uses kernels with variational parameters which can be trained to align the kernel to the ideal kernel for a given dataset \cite{quantum_kernel_alignment_2022}. This could enable the use of a general quantum kernel architecture that can be trained for different datasets.

The remainder of this paper is structured as follows:
The next subsection \ref{Related Work} gives an overview of the related work in the fields of QKMs and QKA.
The final part of the introduction contains our contributions (subsection \ref{Contributions}).
In the following Background (section \ref{Background}), the fundamentals of supervised learning, quantum kernels, and QKA are introduced.
Next, the Methods (section \ref{Methods}) contain the implementation of the model and the Experiments (section \ref{Experiments}) describe the numerical experiments carried out.
In the Results and Discussion (section \ref{Results and Discussion}) we show the results of our experiments and analyze them.
Finally, Conclusion and Outlook (section \ref{Conclusion and Outlook}) highlights the key results of this work and gives future research directions.

\subsection{Related Work} \label{Related Work}
In 2021, Hubregetsen et al. \cite{quantum_kernel_alignment_2022} described the algorithm of QKA and defined the kernel-alignment measure. Moreover, they theoretically assessed the influence of noise on the algorithm and carried out numerical experiments on toy datasets, both on simulations and on real hardware.

Gentinetta et al. \cite{QKA_w_stoch_grad_desc_Gentinett_2023} developed a Quantum Support Vector Machine (QSVM) for which the quantum kernel is trained with QKA using the Pegasos algorithm in 2023. Unlike the default Support Vector Machine (SVM) implementation, their algorithm solves the primal formulation of the SVM which results in a min-min optimization and hence the SVM weights and the kernel parameters can be optimized simultaneously, increasing the efficiency of the algorithm. 

In the same year, Kölle et al. \cite{kölle2023efficient} introduced a one-class QSVM, for which they reduced the training and inference times compared to a default QSVM by up to 95\% and 25\%, respectively, by applying randomized measurements and the variable subsampling ensemble method while achieving a superior average precision compared to a SVM with Radial Basis Function (RBF) kernel.

Finally, in 2024 Bowles et al. \cite{schuld_qml_benchmarks2024} benchmarked 12 popular QML models on six binary classification tasks. They concluded that out-of-the-box classical ML models tend to outperform the QML models and that entanglement does not necessarily improve the models' performance. Moreover, they noted that QML models both in simulations and on hardware can usually only handle input with size of the order of tens of features, and therefore classical pre-processing techniques such as principal component analysis are required to deal with higher dimensional data like the famous MNIST dataset. This classical pre-processing, however, influences the performance of the model and hence dilutes the results of a benchmark.

\subsection{Contributions} \label{Contributions}
Our work aims to answer this question: Can the time complexity of QKMs be improved while still achieving satisfactory results?

We found a positive answer and report these contributions of our work:
\begin{itemize}
    \item We develop Quantum Aligned Centroid Kernel (QUACK), a classifier based on quantum kernel alignment that improves the time complexity compared to basic kernel methods from $\bigO(n^2_\text{train})$ to $\bigO(n_\text{train})$ during training and from $\bigO(n_\text{train}n_\text{test})$ to $\bigO(n_\text{test})$ during testing.
    \item We benchmark our classifier by evaluating it on eight different datasets with up to 784 features and different class ratios ranging from balanced to highly unbalanced.
    \item Finally, we observe that QUACK performs on a similar level as a classical SVM with RBF kernel
\end{itemize}

\section{Background} \label{Background}
\subsection{Supervised Learning} \label{Supervised Learning}
Let $\mathcal{X}\subset\mathbb{R}^d$ be the data space and $\boldsymbol{x} = (x_1,\ldots,x_d) \in \mathcal{X}$ the feature vector of a single $d$-dimensional sample.
Let $\mathcal{Y}$ denote the target variable space and $y \in \mathcal{Y}$ the target variable or label of a single sample. For the case of binary classification, we restrict the target variable to the set $\{1,-1\}$.
Let further $\Theta$ denote the space of model parameters.
The general task of \textit{supervised} machine learning is to train a parameterized model $f_\theta: \mathcal{X} \times \Theta\to\mathcal{Y}$ such that it approximates a mapping between input $\boldsymbol{x}$ and output $\hat{y}$ based on the learned parameters $\theta$, as described in \eqref{eq_output_ad}.
During training, the parameters $\theta$ are optimized such that the loss $\mathcal{L}$  quantifying the difference between the predicted output $\hat{y}$ and the target $y$ is minimized, as in \eqref{eq_loss_ad}.
\begin{align}
    \hat{y} = f_\theta(x) \label{eq_output_ad} \\
    \min_\theta \mathcal{L}(y, \hat{y}) \label{eq_loss_ad}
\end{align}

\subsection{Quantum Kernels} \label{Quantum Kernels}
Quantum kernels emerge as an important tool for encoding classical data into quantum systems and subsequently classifying the data.  It is hoped that the unique properties of quantum computing, such as entanglement and superposition, which are utilized in quantum kernels, will enable them to be more powerful than classical kernels. This hypothesis is supported by theoretical results showing that a constructed classification problem based on the discrete logarithm can be efficiently solved by QKMs, but not by classical ML methods \cite{quantum_kernel_advantage_discrete_log_2021}. 
In general, the encoding in QKMs is achieved through unitary operations $U(\boldsymbol{x}_i)$ that depend on individual data points $\boldsymbol{x}_i$, often implemented via Pauli rotations. The state of the system after the encoding is
\begin{align}
\ket{\psi(\boldsymbol{x}_i)} = \ket{\psi_i} = U(\boldsymbol{x}_i) \ket{0}.
\end{align}

Kernels are known from classical machine learning, where they are real- or complex-valued positive definite functions of two data points, i.e. $\kappa : \mathcal{X} \times \mathcal{X} \rightarrow \mathbb{K}$, where $\mathbb{K} \in \{\mathbb{R}, \mathbb{C}\}$.
This definition can be extended to the quantum case, where a kernel $k$ between two pure data-encoding quantum states $\psi_{i}$ and $\psi_{j}$ is calculated from the fidelity between these states
\begin{align}
     k(\boldsymbol{x}_i, \boldsymbol{x}_j) &= F(\psi_{i}, \psi_{j}) = \vert \braket{\psi_{i}|\psi_{j}} \vert ^2 \label{eq:quantum_kernel_sim} \\
     &= \vert \bra{0^{\otimes n}} U^\dag(\boldsymbol{x}_i) U(\boldsymbol{x}_j) \ket{0^{\otimes n}} \vert ^2 \label{eq:quantum_kernel_hardware}
\end{align}
with data encoding unitaries $U(\boldsymbol{x}_j)$ and $U^\dag(\boldsymbol{x}_i)$, where $U^\dag$ denotes the conjugate transpose of $U$.
This quantum kernel serves as a similarity measure between the states of two encoded samples:
If both samples are identical, i.e. $\boldsymbol{x}_i = \boldsymbol{x}_j$, so $\psi_{i} = \psi_{j}$ as well, the kernel equation \eqref{eq:quantum_kernel_sim} simplifies to
\begin{align}
     k(\boldsymbol{x}_i, \boldsymbol{x}_j) = k(\boldsymbol{x}_i, \boldsymbol{x}_i) = F(\psi_{i}, \psi_{i}) = \vert \braket{\psi_{i}|\psi_{i}} \vert ^2  = 1.
     \label{eq:quantum_kernel_same_state}
\end{align}

On the other hand, if the encoded states $\psi_{i}$ and $\psi_{j}$ are orthogonal, the kernel will evaluate to 
\begin{align}
     k(\boldsymbol{x}_i, \boldsymbol{x}_j) &= F(\psi_{i}, \psi_{j}) = \vert \braket{\psi_{i}|\psi_{j}} \vert ^2 = 0.
     \label{eq:quantum_kernel_orth_state}
\end{align}

\begin{figure}[htbp]
\centerline{\includegraphics[width=.5\textwidth]{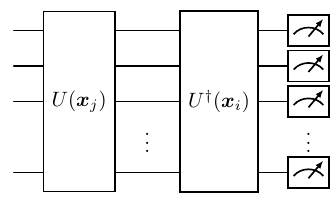}}
\caption{Architecture of the circuit if executed on hardware. The kernel entry $K_{ij}$ for samples $i$ and $j$ is the probability of measuring the all-zero bit string.}
\label{fig:circuit}
\end{figure}

A quantum kernel can be implemented as an $n$-qubit circuit that consists of a trainable unitary $U(\boldsymbol{x}_j)$, encoding a single sample, followed by the complex conjugate $U^\dagger(\boldsymbol{x}_i)$ of another sample, and a measurement of all qubits, as shown in Fig.~\ref{fig:circuit}. The kernel value of the two samples is then obtained as the probability of measuring the all-zero state as given in equation \eqref{eq:quantum_kernel_hardware}. If a state vector simulator is used, the kernel value of two samples $\boldsymbol{x}_i$ and $\boldsymbol{x}_j$ is the fidelity of the states after application of the unitary $U(\boldsymbol{x}_i)$, respectively $U(\boldsymbol{x}_j)$, as given in \eqref{eq:quantum_kernel_sim}.

\subsection{Trainable Quantum Kernels}
A quantum kernel can contain not only parameters that encode the data into the circuit, but also adjustable parameters that affect the performance of the kernel for a particular dataset. This can for example be achieved by alternating layers of rotational gates, whose parameters consist of either one or more features of the datum $\boldsymbol{x}$ or some other variational parameter $\boldsymbol{w}$. These parameters $\boldsymbol{w}$ can be optimized through Quantum Kernel Alignment (QKA), as explained in subsection \ref{Quantum Kernel Alignment}. 
For variational circuits, \textit{trainable encodings} seem to be promising, since they were found to improve the robustness and generalization of the model \cite{training_robust_and_generalizable_qms_2023, trainable_encodings_Jaderberg}. A trainable encoding embeds the datum $\boldsymbol{x}$ and the variational parameter $\boldsymbol{w}$ and bias $\boldsymbol{b}$ as one parameter vector $\boldsymbol{\theta}$, where each entry of $\boldsymbol{\theta}$ is a single parameter used in a rotational gate. The parameter vector $\boldsymbol{\theta}$ is calculated as
\begin{align}
    \boldsymbol{\theta} = \boldsymbol{w}\circ \boldsymbol{x} + \boldsymbol{b},
\end{align}
analog to the neurons of neural networks, where $\circ$ is the element-wise product (Hadamard product).

\subsection{Quantum Kernel Alignment} \label{Quantum Kernel Alignment}
QKA is a powerful tool that can be used to align a trainable kernel to the ideal kernel for a given dataset.  It was originally developed for classical kernels \cite{kernel_alignment_2001} but can be used for quantum kernels as well. 
The implementation of QKA in this work is based on \cite{quantum_kernel_alignment_2022}.
Kernel Alignment is used to optimize the kernel parameters for a specific task, improving the performance of kernel-based algorithms.
The ideal kernel $k^*$ is defined such that it always outputs the correct similarity between two data points:
\begin{align}
k^*\left(\boldsymbol{x}_i, \boldsymbol{x}_j\right)= \begin{cases}1 & \text { if } \boldsymbol{x}_i \text { and } \boldsymbol{x}_j \text { in same class } \\ -1 & \text { if } \boldsymbol{x}_i \text { and } \boldsymbol{x}_j \text { in different classes }\end{cases} \label{eq:ideal_kernel}
\end{align}
In general, this ideal kernel is not known, but for the training set the ideal kernel matrix $K^*$ can be constructed from the labels, i.e. $K^*_{ij} = y_i y_j$, or in vectorized form
\begin{align}
K^* = \boldsymbol{y} \boldsymbol{y}^T.
\end{align}
The kernel-target alignment is a measure of the similarity between two kernels. To calculate it, we need the \textit{Frobenius inner product} between two matrices as defined in \eqref{eq:frobenious_inner_product}.
\begin{align}
\langle A, B\rangle_F=\sum_{i j} A_{i j} B_{i j}=\operatorname{Tr}\left\{A^T B\right\} \label{eq:frobenious_inner_product}
\end{align}
With this, the kernel-target alignment $\mathrm{TA}$ between the current kernel $K$ and the ideal kernel $K^*$ can be calculated as in \eqref{eq:kernel_target_alignment}. 
\begin{align}
\mathrm{TA}(K) &= \frac{\left\langle K, K^*\right\rangle_F}{\sqrt{\langle K, K\rangle_F\left\langle K^*, K^*\right\rangle_F}}\notag \\ &=  \frac{\sum_{i j} y_i y_j k\left(\boldsymbol{x}_i, \boldsymbol{x}_j\right)}{\sqrt{\left(\sum_{i j} k\left(\boldsymbol{x}_i, \boldsymbol{x}_j\right)^2\right)\left(\sum_{i j} y_i^2 y_j^2\right)}}\label{eq:kernel_target_alignment}
\end{align}
The numerator of \eqref{eq:kernel_target_alignment}, $\sum_{i j} y_i y_j k\left(\boldsymbol{x}_i, \boldsymbol{x}_j\right)$, is the \textit{kernel polarity}. If two samples are in the same class, $y_i y_j = 1$, the kernel value $k(\boldsymbol{x}_i, \boldsymbol{x}_j)$ will increase the sum and hence the kernel-target alignment. For samples in different classes, $y_i y_j = -1$, the kernel polarity decreases by $k(\boldsymbol{x}_i, \boldsymbol{x}_j)$ and subsequently the kernel-target alignment decreases, too.
The kernel-target alignment equals 1 if the matrices are perfectly aligned and -1 if they are perfectly misaligned, i.e. perfectly inversely correlated.

\section{Methodology} \label{Methods}
Driven by the need of a NISQ compatible quantum classification algorithm that can handle data on an industrially relevant scale, we developed QUACK, a linear complexity algorithm for supervised classification based on quantum kernel alignment. QUACK is motivated by the desire to find a quantum kernel algorithm that avoids the calculation of the pairwise distances between the samples. Instead, our algorithm optimizes the distance using centroids as a proxy for each class with labels $l\in\{1, -1\}$. The centroids are intitialized as the means of the classes in the original input data space. However, during training of the embedding map, the initial centroids cease to represent the center of the classes and we need to update them. This results in a two step alternating training procedure, where we iteratively optimize the parameters of the embedding map, followed by the position of one of the centroids. Since we do not want to store the centroids as a vector in the $2^n$-dimensional embedding space, we optimize the preimage of the centroids in embedding space, i.e. their positions in data space. Additionally, we alternate the class of the centroid to be optimized in each iteration.

\begin{figure*}[htbp]
\centerline{\includegraphics[width=\textwidth]{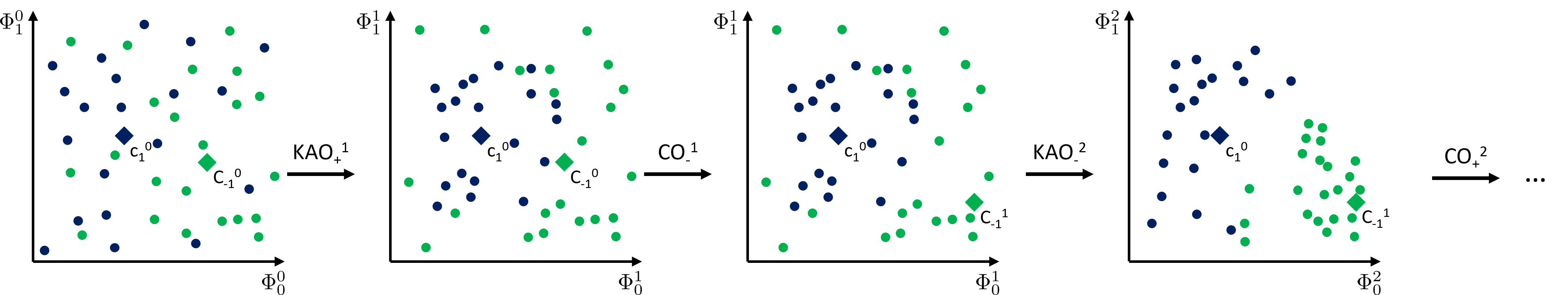}}
\caption{Alternating optimization procedure for QUACK. The blue (green) dots show samples of class 1 (-1) in the embedding space $\Phi$. The superscript defines the current epoch and the subscript the dimension. The diamonds represent the centroids of the classes, where the superscript is the epoch in which the centroid has been optimized the last time and the subscript is the centroid class. For the KAO and CO steps, the superscript gives the epoch and the subscript the class for which the optimization is carried out with $+$ representing class 1 and $-$ representing class -1.}
\label{fig:quack_steps}
\end{figure*}

The working principle of the algorithm is illustrated in Fig.~\ref{fig:quack_steps}. The centroids are initialized as the mean of each class in data space and initially, the embedding map performs a random embedding of the data in Hilbert space, as shown in the first figure. The first step of the algorithm, Kernel Alignment Optimization (KAO) iteration 1 for class 1 optimizes the parameters of the embedding map such that the distances between the samples of class 1 and centroid 1 are minimized, and the distances between the samples of class -1 and centroid 1 are maximized. This results in a new embedding map which is shown in the second figure. Next, the Centroid Optimization (CO) iteration 1 class -1 optimizes the position in data space of the centroid of the other class (class -1) with the aim of minimizing the distances between the centroid and the samples of class -1 and maximizing the distances between centroid -1 and the samples of class 1. The result of the CO step is shown in the third figure. After this, the first epoch of the QUACK algorithm is complete, and a new iteration of the two step process starts. For the second iteration of the KAO, the embedding map is optimized such that the distances between the samples of class -1 and centroid -1 are minimal, and the distances between the samples of class 1 and centroid -1 are maximal. The resulting new embedding space is shown in the fourth figure. Next, the second iteration of the CO is carried out, where the position of centroid 1 is optimized in data space, followed by the next epoch of the two step process and so on.

The parameterized circuit with trainable encoding used for data encoding is described in subsubsection \ref{Circuit Design and Data Encoding} Circuit Design and Data Encoding. The two-step optimization process is explained in subsubsection \ref{Kernel Alignment Optimization} Kernel Alignment Optimization and subsubsection \ref{Centroid Optimization} Centroid Optimization. Finally, the Prediction Stage - where the kernel entries of a new sample with the centroids are estimated and the sample is given the label of the class for whose centroid the kernel entry is maximal - is explained in subsubsection \ref{Prediction Stage}. The final part of this section, subsection \ref{State Vector Simulator} sketches very briefly how our state vector simulator works.

\subsection{QUACK} \label{QUACK}
The QUACK training algorithm is sketched in algorithm \ref{algo:quack} and can be summarized as follows: The algorithm estimates a quantum kernel for the train samples $X$ and the current working centroid $\boldsymbol{c}_l \in \{\boldsymbol{c}_{-1}, \boldsymbol{c}_1\}$. Each of the $n_\text{epochs}$ training epoch consists of a two-step optimization iterating between $n_\text{KAO}$ epochs of optimizing the model parameters $\boldsymbol{w}, \boldsymbol{b}$ and $n_\text{CO}$ epochs of optimizing the centroids $\boldsymbol{c}_{-1}, \boldsymbol{c}_1$. For predicting new data, the kernel values of the new data $X_{\text {predict}}$ and both centroids will be calculated, and each sample is given the label of the centroid with higher kernel entry. In the following, the different parts of the algorithm will be described in more detail.

\begin{algorithm}
\caption{QUACK training} \label{algo:quack}

\algblockdefx{MRepeat}{EndRepeat}[1]{\textbf{Repeat} #1 \textbf{times}:}{}
\algnotext{EndRepeat}

\begin{tikzpicture}
\node (algo) [text width=\linewidth-2cm, inner sep=0pt, anchor=north west] at (0,0) {
\begin{minipage}{\textwidth}
\begin{algorithmic}[1]
\Statex \textbf{Input: }initial guess for $\boldsymbol{c}_{-1}, \boldsymbol{c}_1$
\State $l \gets \text{random bit} \cdot 2 - 1$ 
\MRepeat{$n_\text{epochs}$}
    \MRepeat{$n_\text{KAO}$}
        \State $L_\text{KAO} \gets$ \Call{$\mathcal{L}_\text{KAO}$}{$X, \boldsymbol{y}, \boldsymbol{c}_{l}, \boldsymbol{w}, \boldsymbol{b}$}
        \State optimize model parameters $\boldsymbol{w}, \boldsymbol{b}$
    \EndRepeat
    \State $l \gets -l$ 
    \MRepeat{$n_\text{CO}$}
        \State $L_\text{CO} \gets$ \Call{$\mathcal{L}_\text{CO}$}{$X, \boldsymbol{y}, \boldsymbol{c}_{l}, \boldsymbol{w}, \boldsymbol{b}$}
        \State optimize $\boldsymbol{c}_{l}$ 
    \EndRepeat
\EndRepeat

\end{algorithmic}
\end{minipage}
    };
  
\node (image) [inner sep=0pt, anchor=south west, xshift=-0mm, yshift=0mm] at (algo.south east) {
    \includegraphics[width=2.5cm]{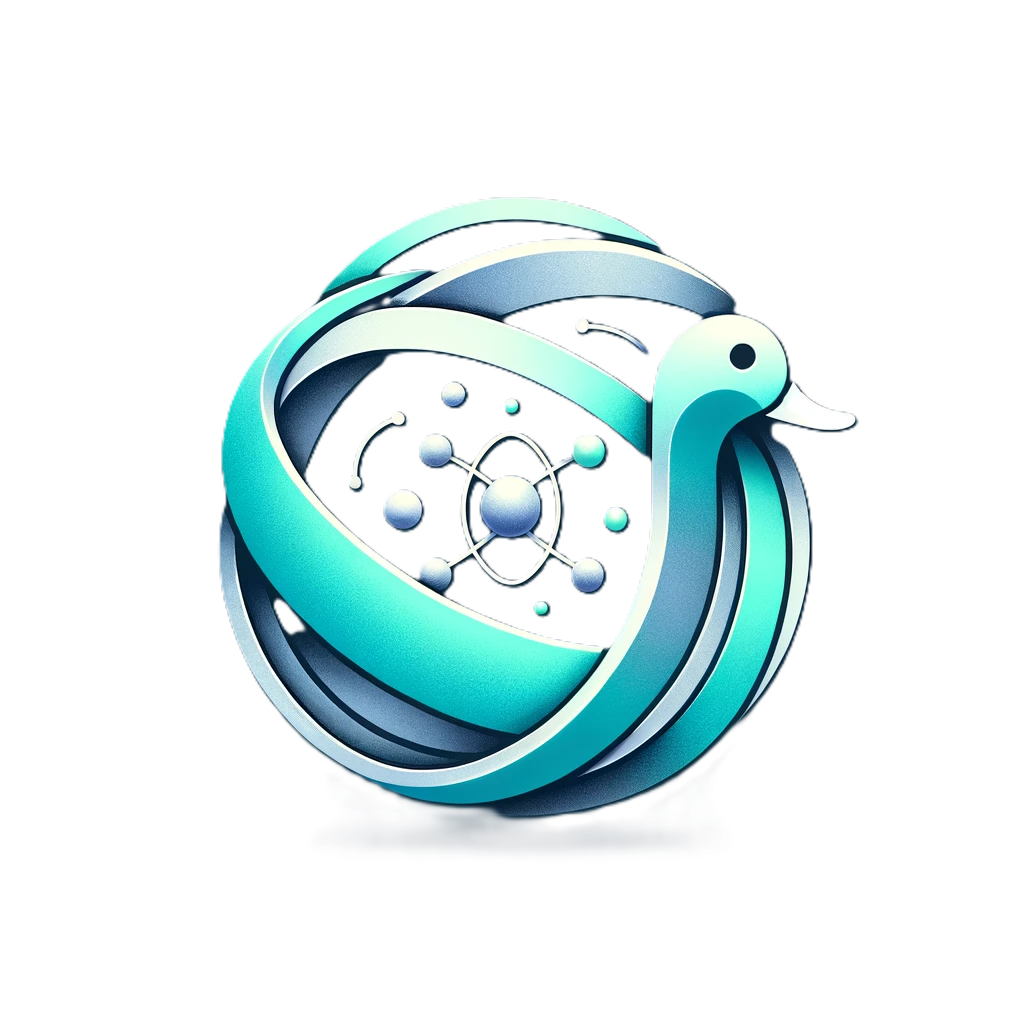}
  };
\end{tikzpicture}

\end{algorithm}

\subsubsection{Circuit Design and Data Encoding} \label{Circuit Design and Data Encoding}
We use a trainable encoding map that was found to yield robustness and generalization improvements over fixed encodings \cite{training_robust_and_generalizable_qms_2023}. In this context, trainable encoding refers to encodings, where the parameters of the gates depend on both, trainable weights and features of the data. How exactly the gate parameters are composed will be defined later.

\begin{figure}[htbp]
\centerline{\includegraphics[width=.5\textwidth]{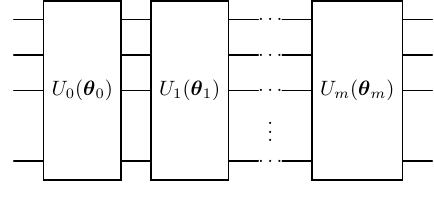}}
\caption{Architecture of the encoding unitary U.}
\label{fig:encoding_unitary}
\end{figure}

\begin{figure}[htbp]
\centerline{\includegraphics[width=.5\textwidth]{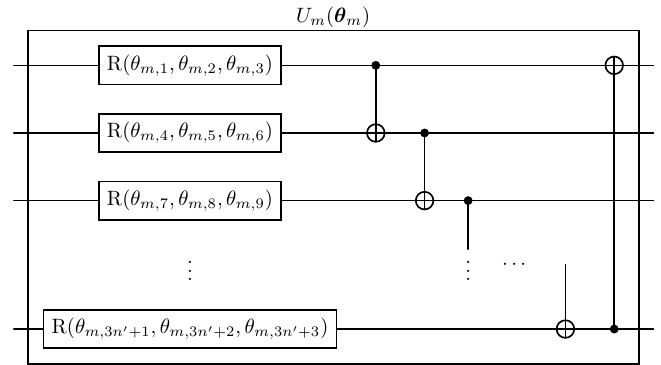}}
\caption{Architecture of the unitary of a single parameterized layer $U_m(\boldsymbol{\theta}_m)$ with $n' = n - 1$.}
\label{fig:one_layer}
\end{figure}

The data encoding unitary $U(\boldsymbol{x}_j)$ consists of $m' = m + 1$ layers of a unitary $U_m(\boldsymbol{\theta}_m)$, as in Fig.~\ref{fig:encoding_unitary}. For clarification, if we have e.g. $m' = 5$ layers, the first layer is $U_0(\boldsymbol{\theta}_0)$ and the last layer is $U_4(\boldsymbol{\theta}_4)$.
Each of the unitaries $U_m(\boldsymbol{\theta}_m)$ is built using a layer of rotation gates and a ring of CNOT-gates, as sketched in Fig.~\ref{fig:one_layer}. The rotation gate is the general parameterized rotation gate \cite{Quantum_Computing_Schuld2021} with matrix representation shown in \eqref{eq:Rot_gate}.
\begin{align}
&R(\theta_{m,i}, \theta_{m,i+1}, \theta_{m,i+2}) = R(\phi, \theta, \omega)=R Z(\omega) R Y(\theta) R Z(\phi)=  \notag\\
&= \left[\begin{array}{cc}
e^{-i(\phi+\omega) / 2} \cos (\theta / 2) & -e^{i(\phi-\omega) / 2} \sin (\theta / 2) \\
e^{-i(\phi-\omega) / 2} \sin (\theta / 2) & e^{i(\phi+\omega) / 2} \cos (\theta / 2)
\end{array}\right] \label{eq:Rot_gate}
\end{align}

Each parameter $\theta_{m,i}$ is calculated from the $k$-th feature of the sample $\boldsymbol{x_j}$, weight $w_{m,i}$ and bias $b_{m,i}$ as given in \eqref{eq:theta_i}, where $k$ is a repeating counter from 1 to the number of input dimensions $d$, i.e. $k = (3nm+ i) \operatorname{mod} d$ for the $n$-qubit system, the $m$-th layer and the $i$-th parameter in the layer.
\begin{align}
    \theta_{m,i} = w_{m,i} \cdot x_{j,k} + b_{m,i} \label{eq:theta_i}
\end{align}

\subsubsection{Kernel Alignment Optimization} 
\label{Kernel Alignment Optimization}

During the Kernel Alignment Optimization, the parameter vectors $\boldsymbol{w}$ and $\boldsymbol{b}$ of the embedding map are optimized. This is achieved by estimating the kernel between the train samples and one centroid and comparing it to the ideal kernel to obtain the kernel alignment. Since we use only one centroid to calculate the kernel, the matrix is an $n_{\text{samples}}$-dimensional vector. 
The ideal kernel is simply the label vector $\boldsymbol{y}$ if the centroid class is 1 and  $-\boldsymbol{y}$ if the centroid class is -1.
The current kernel entries are the fidelities between the current centroid $\boldsymbol{c}_l$ and the samples, where $l \in \{-1,1\}$ defines the class label of the current centroid:
\begin{align}
k(\boldsymbol{c}_l, \boldsymbol{x}_i) = | \braket{\psi_{l} | \psi_i} | ^2 
\end{align}
To get the kernel alignment, equation~\eqref{eq:kernel_target_alignment} is adapted for vectors as kernels and we obtain:
\begin{align}
\mathrm{TA} &=  \frac{l\cdot \sum_{i} y_i k\left(\boldsymbol{c}_l, \boldsymbol{x}_i\right)}{\sqrt{\left(\sum_{i} k\left(\boldsymbol{c}_l, \boldsymbol{x}_i\right)^2\right)\sum_{i} y_i^2}}\label{eq:KAO_kernel_target_alignment}
\end{align}

The loss function $f_{\boldsymbol{c}_l}$ is derived from the kernel-target alignment, with an additional regularization term with regularization parameter $\lambda_{\text{KAO}}$. Note that in this loss function, the centroid $\boldsymbol{c}_l$ is fixed.
\begin{align}
\mathcal{L}_\text{KAO} = f_{\boldsymbol{c}_l}(\boldsymbol{w}, \boldsymbol{b}) = 1 - \mathrm{TA} + \lambda_{\text{KAO}} ||\boldsymbol{w}||^2_2
\end{align}

This loss function is then used to optimize the kernel parameters $\boldsymbol{w}$ and $\boldsymbol{b}$ either through backpropagation if the circuits are executed on a simulator or the parameter shift rule if real hardware is used, by solving this minimization problem:
\begin{align}
    \min_{\boldsymbol{w}, \boldsymbol{b}} f_{\boldsymbol{c}_l}(\boldsymbol{w}, \boldsymbol{b})
\end{align}

\subsubsection{Centroid Optimization} \label{Centroid Optimization}
The Centroid Optimization optimizes the position of the current centroid $\boldsymbol{c}_{l}$ in data space. For this, the kernel alignment is calculated the same way it is in the KAO optimization and then converted to a loss function $g_{\boldsymbol{w}, \boldsymbol{b}}$, in which the parameters $\boldsymbol{w}$ and $\boldsymbol{b}$ of the embedding map are fixed:
\begin{align}
\mathcal{L}_\text{CO} = g_{\boldsymbol{w}, \boldsymbol{b}}(\boldsymbol{c}_l) = 1 - \mathrm{TA} + \lambda_{\text{CO}} R
\end{align}
 Since the features are normalized in the range $(0,1)$, the regularization $R$ introduces a penalty if the centroid position in data space is outside the normalization range:
\begin{align}
R &= \sum_d \left(\max(\boldsymbol{c}^d_{l}-1  ,0) - \min(\boldsymbol{c}^d_{l}, 0)\right),
\end{align}
where $\boldsymbol{c}^d_{l}$ is the $d$-th entry of the vector $\boldsymbol{c}_{l}$. This loss is then used to optimize the position of the working centroid $\boldsymbol{c}_{l}$ in data space, by solving the following minimization problem:
\begin{align}
    \min_{\boldsymbol{c}_l} g_{\boldsymbol{w}, \boldsymbol{b}} (\boldsymbol{c}_l)
\end{align}

\subsubsection{Prediction Stage} \label{Prediction Stage}
After the training is complete, the labels of new samples can be predicted. For this, the kernel $K_\text{predict}$ of shape $(n_\text{samples}, 2)$ between the new samples $X_\text{predict}$ and both centroids $\boldsymbol{c}_{-1}$ and $\boldsymbol{c}_1$ is calculated. Each sample is given a label according to \eqref{eq:y_pred}, where $K_{i,1}$ represents the kernel value for sample $i$ and centroid 1, and $K_{i,-1}$ the value for sample $i$ and centroid -1.
\begin{align}
    \hat{y}_i = \operatorname{sign}(K_{i,1} - K_{i,-1})
    \label{eq:y_pred}
\end{align}

\subsection{State Vector Simulator} \label{State Vector Simulator}
To speed up the simulations, a state vector simulator is implemented using PyTorch \cite{pytorch2019}. The simulator builds up on the \texttt{nn.Module} class where the forward function returns the states of shape $(n_{\text{samples}}, 2^{n_{\text{qubits}}})$ after applying all gates. The main advantage of the simulator is that the unitary of each gate is applied to all $n_{\text{samples}}$ states in one operation, making the execution of the circuits for multiple samples faster when compared to other commonly used simulators.

\section{Experiments} \label{Experiments}
Our linear time complexity QUACK algorithm is benchmarked on eight binary datasets from different areas and with various numbers of features and class ratios. The performance of our model is compared to three other models, containing both classical and quantum approaches.
The source code can be found in a public code repository\footnote{\url{https://github.com/Fraunhofer-AISEC/QUACK/tree/v1}}.

\subsection{Datasets}
The model is benchmarked on eight datasets from different areas, including IT security and handwritten digits. An overview of the datasets is given in table \ref{tab:datasets_overview}. The number of features varies between 14 for the Census dataset and 784 for the image datasets. The share of the smaller class in the total dataset varies between 0.09 (0.07/0.09) and 0.50 (0.49/0.49) for the train (validation/test) set, meaning there are both balanced and highly unbalanced datasets.
For the datasets that have not been pre-split into test and train sets, the train (test) set is created by randomly selecting $70\%$ ($30\%$) of the samples from the dataset. Next, 1,000 samples are randomly selected from the training set for training, 400 from the test set for validation, and a further 400 from the test set for the final testing. 
The class labels are $\{1,-1\}$ according to the criteria specified in table~\ref{tab:datasets_overview}.

\begin{table*}[htbp]
\centering
\caption{Overview of the datasets used in the benchmark. The ratios show the ratio of the minority class to the number of samples in the set.}\label{tab:datasets_overview}
\begin{tabular}{cclccrrrr}
\toprule 
Dataset & Ref. & Description & Class 1 &  Class -1 & Ratio Train & Ratio Val. & Ratio Test & Features \\
\midrule 
Census & \cite{Census} & Income & $\leq50K$ & $>50K$ & 0.28 & 0.26 & 0.23 & 14 \\
CoverT & \cite{CoverT} & Forest tree types & $4$ & $>4$ & 0.09 & 0.07 & 0.09 & 15 \\
DoH & \cite{DoH} & Network traffic & Benign & Malicious & 0.24 & 0.23 & 0.21 & 33 \\
EMNIST & \cite{EMNIST} & Handwritten letters & A-M & N-Z & 0.50 & 0.48 & 0.48 & 784 \\
FMNIST & \cite{FMNIST} & Clothing types & 0-4 & 5-9 & 0.50 & 0.49 & 0.49 & 784 \\
KDD & \cite{KDD} & Network intrusion & Normal & Anomalous & 0.50 & 0.44 & 0.47 & 42 \\
MNIST & \cite{MNIST} & Handwritten digits & 0-4 & 5-9 & 0.48 & 0.47 & 0.49 & 784 \\
URL & \cite{URL} & URLs & Benign & Non-benign & 0.15 & 0.14 & 0.15 & 79 \\
\bottomrule 
\end{tabular}
\end{table*}

\subsection{Models for Benchmarking}

For a comprehensive evaluation of our model, it is benchmarked against these three other approaches: 1) A \textit{vanilla SVM with RBF kernel} and default parameters, implemented with scikit-learn \cite{scikit-learn}. This model has a quadratic time complexity during training and a linear during testing 2) An \textit{RBF centroid classifier}. This classifier first determines the mean of each class in feature space and then calculates a RBF kernel of shape $(n_{\text{samples}}, 2)$ that contains the kernel values between the samples and the means of both classes. Each sample is given the label of the class for which the kernel value between the sample and the respective centroid is larger. This classifier has a linear time complexity and does not require any training. 3) A \textit{Quantum Support Vector Machine} that uses the trained kernel parameters from our approach. However, the QSVM determines the kernels in their quadratic form $(n_{\text{train}}, n_{\text{train}})$ for the training of the SVM parameters and of the shape $(n_{\text{test}}, n_{\text{train}})$ for testing. The SVM hyperparameters are the default ones from scikit-learn.

\subsection{Implementation Details}
All models are trained three times with different random seeds and the mean results with standard deviation are reported.
The hyperparameters used for the QUACK algorithm are shown in  tables~\ref{app:tab:params} and \ref{app:tab:params_shared} in Appendix \ref{app:hyperparameters}. The hyperparameter tuning is achieved by a randomized grid search for each dataset where the validation set is used to evaluate the model. The number of layers and qubits were selected such that each feature is encoded at least once and the model can still be simulated in a reasonable time. 

\subsection{Verification of the Simulator}
To make sure that the state vector simulator works as intended, a small model is trained on both our simulator and PennyLane's \cite{pennylane} \textit{default.qubit} simulator with identical hyperparameters. After training, the weights, biases, loss, and metrics of both models were identical, and therefore we can conclude that our simulator works as intended.

\section{Results and Discussion} \label{Results and Discussion}
The newly introduced linear time complexity algorithm QUACK is benchmarked together with three other models on eight binary datasets from different areas with various numbers of features and class ratios. Each model is run three times on each dataset with different random seeds.

\subsection{Model Performance}
\begin{figure*}[htb]
\centerline{\includegraphics[width=\textwidth]{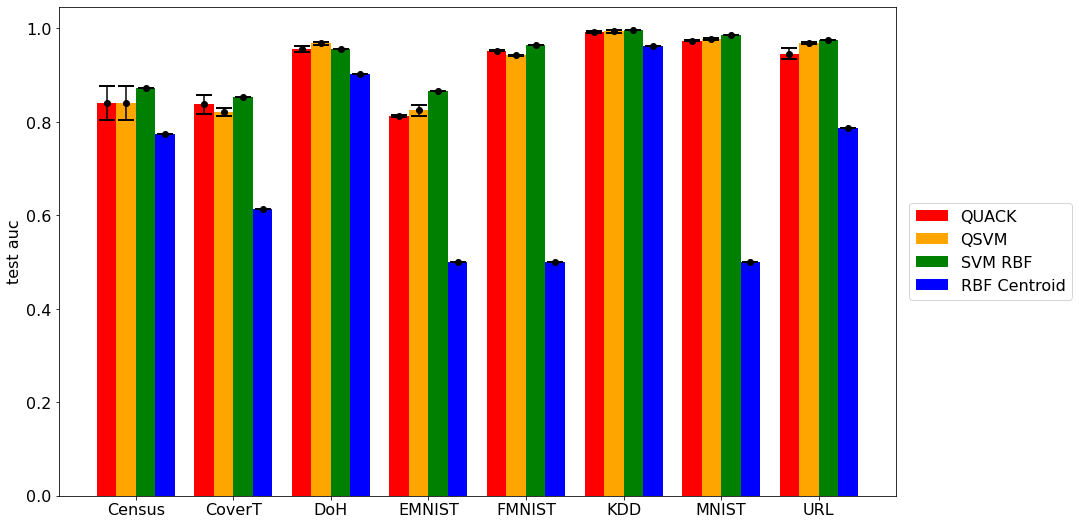}}
\caption{Test AUCs of the different models.}
\label{fig:test_aucs}
\end{figure*}

\begin{figure}[H]
\centerline{\includegraphics[width=\textwidth]{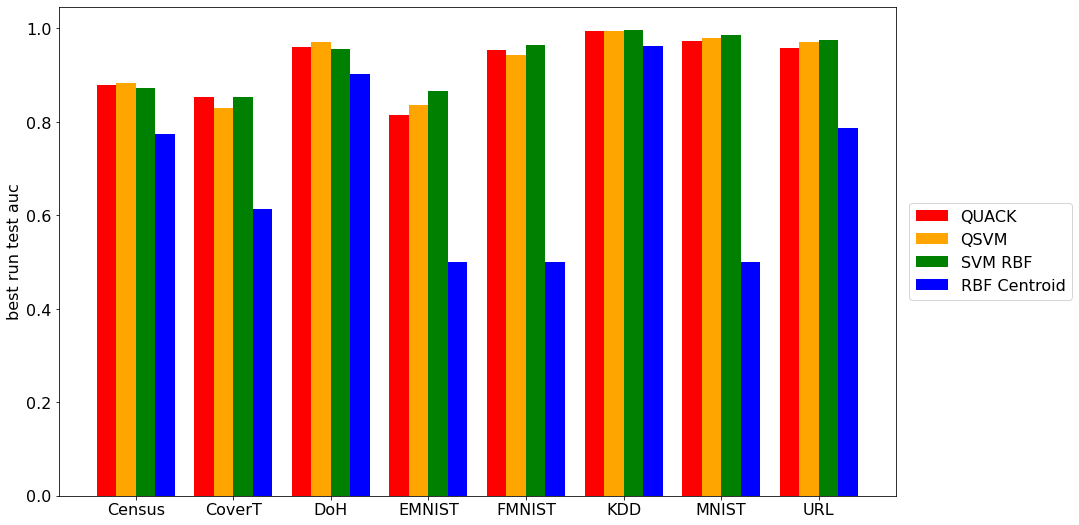}}
\caption{Test AUCs of the best run of each model for each dataset.}
\label{app:fig:test_aucs}
\end{figure}

Figure~\ref{fig:test_aucs} shows the average of the test area under the ROC curve (AUC) for each model and dataset, and the values are listed in table~\ref{app:tab:all_metrics} in Appendix \ref{app:detailed_results}.
Our model performs equally or almost equally as the SVM RBF on five out of eight datasets. More precisely, for CoverT, DoH, FMNIST, KDD, and MNIST, the difference in AUC is 0.02 at most. The performance gap is highest on EMNIST with an AUC difference of 0.06, followed by Census and URL with 0.03. Fig.~\ref{app:fig:test_aucs} shows the test AUCs of the best run of each model. The main difference to Fig.~\ref{fig:test_aucs} is that the best QUACK run additionally achieves equal results as the SVM on the Census dataset. From this, we conclude that QUACK performs on a similar level as the SVM and may be a reasonable alternative to the SVM.

The QSVM that uses the trained weights from our model to compute the full kernel, achieves very similar AUCs compared to our model, with the highest difference being 0.02 in both directions. This suggests, that once the kernel training is completed and the kernel parameters are set, the SVM training and inference methods do not notably improve the model's performance. 

The RBF centroid classifier is the worst model on all datasets, which is intuitive since this model does not require any training at all. It is surprising, however, that this classifier comes relatively close to the performance of the other models for the DoH and KDD datasets. Together with the particularly good performance of the other models on these two datasets, we suspect that DoH and KDD are relatively easy datasets for binary classification. This observation is consistent with the findings of other authors which also reported high AUCs on these two datasets \cite{Schulze2022, Schulze2022_2}.

Finally, the observed standard deviation for QUACK across datasets is consistently low, being 0.01 or below, except for Census and CoverT. These two datasets are notable outliers with a standard deviation of 0.04 and 0.02, respectively. From this we conclude that QUACK's performance is largely independent of the initialisation of the trainable parameters, but is not entirely stable. The QSVM shows a similar standard deviation as QUACK which was expected since both algorithms use the same optimized weights and biases. The SVM RBF and RBF Centroid exhibit no standard deviation, as they are deterministic algorithms.

\subsection{Number of Circuit Evaluations}
\begin{figure}[htb]
\centerline{\includegraphics[width=\linewidth]{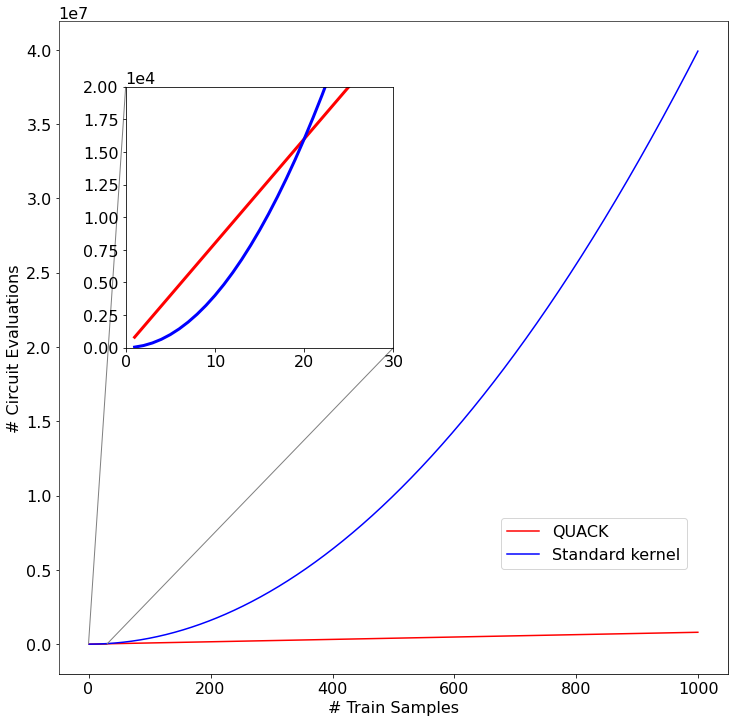}}
\caption{Comparison of the number of evaluation steps between QUACK and a standard kernel. The inset shows a zoom-in of the plot for the number of train samples in the range from 0 to 30.}
\label{fig:num_circ_evals}
\end{figure}

We compare the number of circuit evaluations required during the training of our model with a standard kernel method. All circuit evaluations that result from an evaluation of the models during training are ignored. Fig.~\ref{fig:num_circ_evals} shows the number of circuit evaluations over the number of train samples for the number of epochs used throughout the numerical experiments (see table~\ref{app:tab:params_shared} in Appendix \ref{app:hyperparameters}). 
The QUACK algorithm scales linearly with the number of training samples $n_\text{train}$, and the number of circuit evaluations is $N_\text{QUACK} = n_\text{epochs} \cdot (n_{\text{KAO}} + n_{\text{CO}}) \cdot n_\text{train}$, where $n_\text{epochs}$ is the number of two-step iterations performed, $n_\text{KAO}$ and $n_\text{CO}$ are the numbers of the Kernel Alignment Optimization steps and Centroid Optimization steps, respectively. A standard kernel on the other hand, requires a quadratic number of circuit evaluations $N_\text{standard kernel} = n_\text{epochs} \cdot n^2_\text{train}$. As soon as the number of samples exceeds the sum of the number of epochs for kernel alignment and centroid optimization, i.e. $n_\text{train} > n_{\text{KAO}} + n_{\text{CO}}$, QUACK requires fewer circuit evaluations than the default kernel. With a further increase in the sample size, the number of circuit evaluations required for QUACK grows quadratically slower than for the standard kernel.

\section{Conclusion and Outlook} \label{Conclusion and Outlook}
We developed QUACK, a classifier based on quantum kernel alignment that improves the time complexity compared to basic kernel methods from $\bigO(n^2_\text{train})$ to $\bigO(n_\text{train})$ during training and from $\bigO(n_\text{train}n_\text{test})$ to $\bigO(n_\text{test})$ during testing. QUACK's training time complexity is a polynomial improvement compared to the SVM. The algorithm was benchmarked by evaluating it on eight different datasets with up to 784 features and various class ratios ranging from balanced to highly unbalanced. The performance was compared to a vanilla SVM with an RBF kernel, an RBF centroid classifier, and a QSVM. We conclude that QUACK performs on a similar level as the classical SVM and that the training of the SVM parameters does not improve the predictions of the model once the kernel parameters are trained. Finally, our algorithm works on data with up to 784 features without dimensionality reduction, which is often required for other state-of-the-art QML models. 

Thanks to the linear scaling of QUACK, the algorithm can be used as quick baseline for future classification tasks: If the performance of QUACK is satisfactory, using more costly classification algorithms does not offer an advantage. If, however, QUACK performs poorly, the application of more costly algorithms, e.g. the quadratic-scaling (Q)SVM, should be considered. Furthermore, there is an intuitive explanation for the classification results of the algorithm: If QUACK performs well, it has found an encoding circuit and centroids which separate the classes into different clusters around these centroids in Hilbert space.

On the other hand, QUACK is limited by the assumption that there exists an embedding in which each class forms a cluster around a different centroid. Finally, the performance of QUACK is dependent on the choice of hyperparameters, and a extensive tuning of these hyperparameters is strongly recommended. 

This work is only a first effort toward increasing the potential of QKMs and the next logical step is to benchmark QUACK on hardware. Further work can extend the algorithm to perform multi-class classification by using $k$ centroids for $k$ classes and running a one-versus-all classification in each iteration of the algorithm. In addition, the stability of the algorithm should be improved to make its performance less dependent on the choice of hyperparameters.

\section*{Acknowledgment}
This research is part of the Munich Quantum Valley, which is supported by the Bavarian state government with funds from the Hightech Agenda Bayern Plus.

\printbibliography

@article{Schulze2022_2,
   author = {Jan-Philipp Schulze and Philip Sperl and Ana Răduţoiu and Carla Sagebiel and Konstantin Böttinger},
   month = {6},
   title = {R2-AD2: Detecting Anomalies by Analysing the Raw Gradient},
   url = {http://arxiv.org/abs/2206.10259},
   year = {2022},
}

@inproceedings{Schulze2022,
   author = {Jan Philipp Schulze and Philip Sperl and Konstantin Bottinger},
   doi = {10.1109/IJCNN55064.2022.9892896},
   isbn = {9781728186719},
   booktitle = {Proceedings of the International Joint Conference on Neural Networks},
   publisher = {Institute of Electrical and Electronics Engineers Inc.},
   title = {Double-Adversarial Activation Anomaly Detection: Adversarial Autoencoders are Anomaly Generators},
   volume = {2022-July},
   year = {2022},
}

@article{trainable_encodings_Jaderberg,
  title = {Let quantum neural networks choose their own frequencies},
  author = {Jaderberg, Ben and Gentile, Antonio A. and Berrada, Youssef Achari and Shishenina, Elvira and Elfving, Vincent E.},
  journal = {Phys. Rev. A},
  volume = {109},
  issue = {4},
  pages = {042421},
  numpages = {10},
  year = {2024},
  publisher = {American Physical Society},
  doi = {10.1103/PhysRevA.109.042421},
  url = {https://link.aps.org/doi/10.1103/PhysRevA.109.042421}
}

@misc{sherbrooke,
  title = {ibm\_sherbrooke},
  howpublished = {\url{https://quantum.ibm.com/services/resources?system=ibm_sherbrooke}},
  note = {Accessed: 2024-03-28}
}

@misc{schuld_qml_benchmarks2024,
      title={Better than classical? The subtle art of benchmarking quantum machine learning models}, 
      author={Joseph Bowles and Shahnawaz Ahmed and Maria Schuld},
      year={2024},
      eprint={2403.07059},
      archivePrefix={arXiv},
      primaryClass={quant-ph}
}

@article{scikit-learn,
  title={Scikit-learn: Machine Learning in {P}ython},
  author={Pedregosa, F. and Varoquaux, G. and Gramfort, A. and Michel, V.
          and Thirion, B. and Grisel, O. and Blondel, M. and Prettenhofer, P.
          and Weiss, R. and Dubourg, V. and Vanderplas, J. and Passos, A. and
          Cournapeau, D. and Brucher, M. and Perrot, M. and Duchesnay, E.},
  journal={Journal of Machine Learning Research},
  volume={12},
  pages={2825--2830},
  year={2011}
}

@misc{pennylane,
      title={PennyLane: Automatic differentiation of hybrid quantum-classical computations}, 
      author={Ville Bergholm and Josh Izaac and Maria Schuld and Christian Gogolin and Shahnawaz Ahmed and Vishnu Ajith and M. Sohaib Alam and Guillermo Alonso-Linaje and B. AkashNarayanan and Ali Asadi and Juan Miguel Arrazola and Utkarsh Azad and Sam Banning and Carsten Blank and Thomas R Bromley and Benjamin A. Cordier and Jack Ceroni and Alain Delgado and Olivia Di Matteo and Amintor Dusko and Tanya Garg and Diego Guala and Anthony Hayes and Ryan Hill and Aroosa Ijaz and Theodor Isacsson and David Ittah and Soran Jahangiri and Prateek Jain and Edward Jiang and Ankit Khandelwal and Korbinian Kottmann and Robert A. Lang and Christina Lee and Thomas Loke and Angus Lowe and Keri McKiernan and Johannes Jakob Meyer and J. A. Montañez-Barrera and Romain Moyard and Zeyue Niu and Lee James O'Riordan and Steven Oud and Ashish Panigrahi and Chae-Yeun Park and Daniel Polatajko and Nicolás Quesada and Chase Roberts and Nahum Sá and Isidor Schoch and Borun Shi and Shuli Shu and Sukin Sim and Arshpreet Singh and Ingrid Strandberg and Jay Soni and Antal Száva and Slimane Thabet and Rodrigo A. Vargas-Hernández and Trevor Vincent and Nicola Vitucci and Maurice Weber and David Wierichs and Roeland Wiersema and Moritz Willmann and Vincent Wong and Shaoming Zhang and Nathan Killoran},
      year={2022},
      eprint={1811.04968},
      archivePrefix={arXiv},
      primaryClass={quant-ph}
}

@misc{kölle2023efficient,
      title={Towards Efficient Quantum Anomaly Detection: One-Class SVMs using Variable Subsampling and Randomized Measurements}, 
      author={Michael Kölle and Afrae Ahouzi and Pascal Debus and Robert Müller and Danielle Schuman and Claudia Linnhoff-Popien},
      year={2023},
      eprint={2312.09174},
      archivePrefix={arXiv},
      primaryClass={quant-ph}
}

@misc{training_robust_and_generalizable_qms_2023,
      title={Training robust and generalizable quantum models}, 
      author={Julian Berberich and Daniel Fink and Daniel Pranjić and Christian Tutschku and Christian Holm},
      year={2023},
      eprint={2311.11871},
      archivePrefix={arXiv},
      primaryClass={quant-ph}
}

@Inbook{Quantum_Computing_Schuld2021,
author="Schuld, Maria
and Petruccione, Francesco",
title="Quantum Computing",
bookTitle="Machine Learning with Quantum Computers",
year="2021",
publisher="Springer International Publishing",
address="Cham",
pages="79--146",
isbn="978-3-030-83098-4",
doi="10.1007/978-3-030-83098-4_3",
url="https://doi.org/10.1007/978-3-030-83098-4_3"
}

@InProceedings{URL,
author="Mamun, Mohammad Saiful Islam
and Rathore, Mohammad Ahmad
and Lashkari, Arash Habibi
and Stakhanova, Natalia
and Ghorbani, Ali A.",
editor="Chen, Jiageng
and Piuri, Vincenzo
and Su, Chunhua
and Yung, Moti",
title="Detecting Malicious URLs Using Lexical Analysis",
booktitle="Network and System Security",
year="2016",
publisher="Springer International Publishing",
address="Cham",
pages="467--482",
isbn="978-3-319-46298-1"
}

@ARTICLE{MNIST,
  author={Lecun, Y. and Bottou, L. and Bengio, Y. and Haffner, P.},
  journal={Proceedings of the IEEE}, 
  title={Gradient-based learning applied to document recognition}, 
  year={1998},
  volume={86},
  number={11},
  pages={2278-2324},
  doi={10.1109/5.726791}}

@INPROCEEDINGS{KDD,
  author={Tavallaee, Mahbod and Bagheri, Ebrahim and Lu, Wei and Ghorbani, Ali A.},
  booktitle={2009 IEEE Symposium on Computational Intelligence for Security and Defense Applications}, 
  title={A detailed analysis of the KDD CUP 99 data set}, 
  year={2009},
  volume={},
  number={},
  pages={1-6},
  doi={10.1109/CISDA.2009.5356528}}

@misc{FMNIST,
      title={Fashion-MNIST: a Novel Image Dataset for Benchmarking Machine Learning Algorithms}, 
      author={Han Xiao and Kashif Rasul and Roland Vollgraf},
      year={2017},
      eprint={1708.07747},
      archivePrefix={arXiv},
      primaryClass={cs.LG}
}

@INPROCEEDINGS{EMNIST,
  author={Cohen, Gregory and Afshar, Saeed and Tapson, Jonathan and van Schaik, André},
  booktitle={2017 International Joint Conference on Neural Networks (IJCNN)}, 
  title={EMNIST: Extending MNIST to handwritten letters}, 
  year={2017},
  volume={},
  number={},
  pages={2921-2926},
  doi={10.1109/IJCNN.2017.7966217}}

@INPROCEEDINGS{DoH,
  author={MontazeriShatoori, Mohammadreza and Davidson, Logan and Kaur, Gurdip and Habibi Lashkari, Arash},
  booktitle={2020 IEEE Intl Conf on Dependable, Autonomic and Secure Computing, Intl Conf on Pervasive Intelligence and Computing, Intl Conf on Cloud and Big Data Computing, Intl Conf on Cyber Science and Technology Congress (DASC/PiCom/CBDCom/CyberSciTech)}, 
  title={Detection of DoH Tunnels using Time-series Classification of Encrypted Traffic}, 
  year={2020},
  volume={},
  number={},
  pages={63-70},
  doi={10.1109/DASC-PICom-CBDCom-CyberSciTech49142.2020.00026}}

@article{CoverT,
title = {Comparative accuracies of artificial neural networks and discriminant analysis in predicting forest cover types from cartographic variables},
journal = {Computers and Electronics in Agriculture},
volume = {24},
number = {3},
pages = {131-151},
year = {1999},
issn = {0168-1699},
doi = {https://doi.org/10.1016/S0168-1699(99)00046-0},
url = {https://www.sciencedirect.com/science/article/pii/S0168169999000460},
author = {Jock A. Blackard and Denis J. Dean},
}

@misc{Census,
  author       = {Kohavi,Ron},
  title        = {{Census Income}},
  year         = {1996},
  howpublished = {UCI Machine Learning Repository},
  note         = {{DOI}: https://doi.org/10.24432/C5GP7S}
}

@InProceedings{URL_dataset_2016,
author="Mamun, Mohammad Saiful Islam
and Rathore, Mohammad Ahmad
and Lashkari, Arash Habibi
and Stakhanova, Natalia
and Ghorbani, Ali A.",
editor="Chen, Jiageng
and Piuri, Vincenzo
and Su, Chunhua
and Yung, Moti",
title="Detecting Malicious URLs Using Lexical Analysis",
booktitle="Network and System Security",
year="2016",
publisher="Springer International Publishing",
address="Cham",
pages="467--482",
isbn="978-3-319-46298-1"
}

@inproceedings{QKA_w_stoch_grad_desc_Gentinett_2023,
   title={Quantum Kernel Alignment with Stochastic Gradient Descent},
   url={http://dx.doi.org/10.1109/QCE57702.2023.00036},
   DOI={10.1109/qce57702.2023.00036},
   booktitle={2023 IEEE International Conference on Quantum Computing and Engineering (QCE)},
   publisher={IEEE},
   author={Gentinetta, Gian and Sutter, David and Zoufal, Christa and Fuller, Bryce and Woerner, Stefan},
   year={2023},
   month=sep }

@article{quantum_kernel_advantage_discrete_log_2021,
   author = {Yunchao Liu and Srinivasan Arunachalam and Kristan Temme},
   doi = {10.1038/s41567-021-01287-z},
   issn = {17452481},
   issue = {9},
   journal = {Nature Physics},
   month = {9},
   pages = {1013-1017},
   publisher = {Nature Research},
   title = {A rigorous and robust quantum speed-up in supervised machine learning},
   volume = {17},
   year = {2021},
}

@misc{supervised_qml_are_kernel_2021,
      title={Supervised quantum machine learning models are kernel methods}, 
      author={Maria Schuld},
      year={2021},
      eprint={2101.11020},
      archivePrefix={arXiv},
      primaryClass={quant-ph}
}

@misc{pytorch2019,
      title={PyTorch: An Imperative Style, High-Performance Deep Learning Library}, 
      author={Adam Paszke and Sam Gross and Francisco Massa and Adam Lerer and James Bradbury and Gregory Chanan and Trevor Killeen and Zeming Lin and Natalia Gimelshein and Luca Antiga and Alban Desmaison and Andreas Köpf and Edward Yang and Zach DeVito and Martin Raison and Alykhan Tejani and Sasank Chilamkurthy and Benoit Steiner and Lu Fang and Junjie Bai and Soumith Chintala},
      year={2019},
      eprint={1912.01703},
      archivePrefix={arXiv},
      primaryClass={cs.LG}
}

@article{quantum_kernel_alignment_2022,
  title = {Training quantum embedding kernels on near-term quantum computers},
  author = {Hubregtsen, Thomas and Wierichs, David and Gil-Fuster, Elies and Derks, Peter-Jan H. S. and Faehrmann, Paul K. and Meyer, Johannes Jakob},
  journal = {Phys. Rev. A},
  volume = {106},
  issue = {4},
  pages = {042431},
  numpages = {18},
  year = {2022},
  publisher = {American Physical Society},
  doi = {10.1103/PhysRevA.106.042431},
  url = {https://link.aps.org/doi/10.1103/PhysRevA.106.042431}
}

@inproceedings{kernel_alignment_2001,
 author = {Cristianini, Nello and Shawe-Taylor, John and Elisseeff, Andr\'{e} and Kandola, Jaz},
 booktitle = {Advances in Neural Information Processing Systems},
 editor = {T. Dietterich and S. Becker and Z. Ghahramani},
 pages = {},
 publisher = {MIT Press},
 title = {On Kernel-Target Alignment},
 url = {https://proceedings.neurips.cc/paper_files/paper/2001/file/1f71e393b3809197ed66df836fe833e5-Paper.pdf},
 volume = {14},
 year = {2001}
}
\clearpage
\onecolumn
\appendices
\section{Hyperparameters}  \label{app:hyperparameters}

\begin{table}[H]
\centering
\caption{overview of the optimized hyperparameters for each dataset.}
\label{app:tab:params}
\begin{tabular}{crrrrrr}
\toprule 
Datasets & $lr_{kao}$ & $lr_{co}$ & $r_\text{decay}$ & $reg_{kao}$ & $reg_{co}$ \\
\midrule 
Census & 0.5 & 0.5 & 0.9 & 0.001 & 0.001 \\
CoverT & 0.5 & 0.1 & 0.9 & 0.001 & 0.001 \\
DoH & 0.5 & 0.5 & 0.9 & 0.001 & 0.001 \\
EMNIST & 1.0 & 5.0 & 0.9 & 0.001 & 0.001 \\
FMNIST & 5.0 & 0.5 & 0.8 & 0.0001 & 0.001 \\
KDD & 0.5 & 1.0 & 0.9 & 0.001 & 0.001 \\
MNIST & 5.0 & 1.0 & 0.9 & 0.001 & 0.001 \\
URL & 0.5 & 0.5 & 0.9 & 0.001 & 0.001 \\
\bottomrule 
\end{tabular}
\end{table}

\begin{table}[h]
\centering
\caption{overview of the hyperparameters shared between QUACK on all datasets. The number of epochs for the two-step training, kernel alignment optimization and centroid optimization are given by $n$, $n_{kao}$, and $n_{co}$ respectively. init\_weights\_scale gives the maximum value for the weights during random initialization.}
\label{app:tab:params_shared}
\begin{tabular}{rrrrrrrrrr}
\toprule 
layers & qubits & $n_\text{train}$ & $n_\text{val}$ & $n_\text{test}$ & $n$ & $n_{kao}$ & $n_{co}$ & init\_weights\_scale & seeds \\
\midrule 
53 & 5 & 1000 & 400 & 400 & 40 & 10 & 10 & 0.1 & 42, 123, 1234 \\
\bottomrule 
\end{tabular}
\end{table}

\section{Detailed Results} \label{app:detailed_results}

\begin{table}[htb]
\centering
\caption{overview of the AUCs of the models.}
\label{app:tab:all_metrics}
\begin{tabular}{crrrrrr}
\toprule
Dataset & train\_auc & val\_auc & test\_auc & qsvm\_train\_auc & qsvm\_val\_auc & qsvm\_test\_auc \\
\midrule
Census & 0.91 ± 0.03 & 0.85 ± 0.02 & 0.84 ± 0.04 & 0.91 ± 0.03 & 0.85 ± 0.02 & 0.84 ± 0.04 \\
CoverT & 0.85 ± 0.03 & 0.75 ± 0.04 & 0.84 ± 0.02 & 0.88 ± 0.02 & 0.73 ± 0.05 & 0.82 ± 0.01 \\
DoH & 0.98 ± 0.00 & 0.96 ± 0.00 & 0.96 ± 0.01 & 0.98 ± 0.00 & 0.98 ± 0.00 & 0.97 ± 0.00 \\
EMNIST & 0.99 ± 0.01 & 0.84 ± 0.01 & 0.81 ± 0.00 & 0.99 ± 0.00 & 0.84 ± 0.01 & 0.82 ± 0.01 \\
FMNIST & 0.98 ± 0.00 & 0.96 ± 0.00 & 0.95 ± 0.00 & 0.99 ± 0.00 & 0.95 ± 0.01 & 0.94 ± 0.00 \\
KDD & 1.00 ± 0.00 & 0.99 ± 0.00 & 0.99 ± 0.00 & 1.00 ± 0.00 & 0.99 ± 0.00 & 0.99 ± 0.00 \\
MNIST & 0.99 ± 0.00 & 0.95 ± 0.01 & 0.97 ± 0.00 & 1.00 ± 0.00 & 0.96 ± 0.01 & 0.98 ± 0.00 \\
URL & 0.93 ± 0.01 & 0.89 ± 0.02 & 0.95 ± 0.01 & 0.95 ± 0.01 & 0.92 ± 0.01 & 0.97 ± 0.00 \\
\bottomrule
\end{tabular}
\bigskip

\begin{tabular}{crrrrr}
\toprule
Dataset & svm\_rbf\_train\_auc & svm\_rbf\_val\_auc & svm\_rbf\_test\_auc & rbf\_centroid\_val\_auc & rbf\_centroid\_test\_auc \\
\midrule
Census & 0.88 ± 0.00 & 0.87 ± 0.00 & 0.87 ± 0.00 & 0.73 ± 0.00 & 0.77 ± 0.00 \\
CoverT & 0.92 ± 0.00 & 0.79 ± 0.00 & 0.85 ± 0.00 & 0.63 ± 0.00 & 0.61 ± 0.00 \\
DoH & 0.98 ± 0.00 & 0.96 ± 0.00 & 0.96 ± 0.00 & 0.90 ± 0.00 & 0.90 ± 0.00 \\
EMNIST & 0.99 ± 0.00 & 0.89 ± 0.00 & 0.87 ± 0.00 & 0.50 ± 0.00 & 0.50 ± 0.00 \\
FMNIST & 0.99 ± 0.00 & 0.97 ± 0.00 & 0.97 ± 0.00 & 0.50 ± 0.00 & 0.50 ± 0.00 \\
KDD & 1.00 ± 0.00 & 1.00 ± 0.00 & 1.00 ± 0.00 & 0.95 ± 0.00 & 0.96 ± 0.00 \\
MNIST & 1.00 ± 0.00 & 0.98 ± 0.00 & 0.98 ± 0.00 & 0.50 ± 0.00 & 0.50 ± 0.00 \\
URL & 0.97 ± 0.00 & 0.95 ± 0.00 & 0.98 ± 0.00 & 0.71 ± 0.00 & 0.79 ± 0.00 \\
\bottomrule
\end{tabular}
\bigskip

\end{table}

\end{document}